# Novel multifunctional plasmonic fiber probe: Enabling plasmonic heating and SERS sensing for biomedical applications

Muhammad Fayyaz Kashif[1,2*], Di Zheng[1,3], Linda Piscopo[1,4], Liam Collard[1], Antonio Balena[1,5], Huatian Hu[1], Daniele Riccio[2], Francesco Tantussi[5], Francesco De Angelis[5], Massimo de Vittorio[1,6§], Ferruccio Pisanello[1§]

[1]Istituto Italiano di Tecnologia, Center for Biomolecular Nanotechnologies, 73010 Arnesano, Italy
[2]Dipartimento di Ingegneria Elettrica e delle Tecnologie dell'Informazione, Università Degli Studi di Napoli Federico II, 80125 Napoli, Italy
[3]State Key Laboratory of Radio Frequency Heterogeneous Integration, Shenzhen University, Shenzhen 518060, China
[4]Dipartimento di Ingegneria dell'Innovazione, Università del Salento, 73100 Lecce, Italy
[5]Laboratoire Kastler Brossel, Sorbonne University, CNRS, ENS-PSL University, Collège de France, Paris, France
[6]Istituto Italiano di Tecnologia, Plasmon Nanotechnologies, Genova 16163, Italy
[7]Department of Health Technology, Drug Delivery and Sensing Section (IDUN), Technical University of Denmark (DTU), Denmark

*Corresponding author. E-mail: muhamadfayyaz.kashif@unina.it, ferruccio.pisanello@iit.it

§ equally contributed and co-last authors

## Abstract

Optical fiber-based platforms are increasingly explored as compact, minimally invasive tools for integrated photonic functionalities in biomedical applications. Among these, the combination of plasmonic heating and optical sensing on a single fiber tip offers compelling opportunities for localized photothermal actuation and *in situ* molecular detection. In this work, we present a multifunctional plasmonic fiber probe (PFP) that enables spectral multiplexing of thermo-plasmonic heating and surface-enhanced Raman spectroscopy (SERS). This dual capability is achieved by integrating gold nanoislands (AuNIs) onto the flat facet of a multimode optical fiber using a solid-state dewetting process—a straightforward and scalable fabrication method that avoids the complexity of lithographic techniques. We characterize how the morphology of the AuNIs modulates optical extinction, photothermal response, and electromagnetic field enhancement across the visible and near-infrared spectrum. Specifically, we demonstrate efficient, wavelength-dependent heating under visible light and strong SERS signal enhancement under near-infrared excitation, both supported by electromagnetic and thermal simulations. The ability to decouple photothermal stimulation and Raman sensing in a single, fiber-integrated device addresses a current gap in lab-on-fiber technologies, where multifunctional operation is often constrained to a single wavelength.

## 1  Introduction

Plasmonic nanoparticles exhibit localized surface plasmon resonances (LSPRs), where light energy at specific wavelengths drives collective oscillations of conduction electrons. This energy can be dissipated via non-radiative decay (absorption), re-emitted through radiative decay (scattering), or remain temporarily confined in intense, non-propagating near fields surrounding the nanoparticle, an effect that underpins many plasmon-enhanced light–matter interactions. These distinct energy channels make plasmonic systems particularly attractive for applications in photothermal heat generation and highly sensitive optical sensing. In the non-radiative channel, plasmon decay proceeds via electron–electron and electron–phonon interactions [1], leading to efficient photothermal conversion and resulting in strong optical absorption and localized heating. This process produces spatially and temporally confined temperature gradients, even under moderate optical excitation [2]. The thermal response can be finely tuned by controlling the nanoparticle's size, shape, material, spatial density, and the excitation wavelength [3], [4],

[5], [6]. On a larger scale, collective heating effects can emerge in dense nanoparticle assemblies due to interparticle plasmonic coupling [7], [8],[9]. In contrast, the plasmon can radiatively decay, re-emitting photons via elastic scattering, whose spectral profile mirrors the resonance. This scattering is strongly dependent on particle size, morphology, and dielectric surroundings, and underlies techniques such as dark-field imaging[10] , colorimetric sensing [11], and plasmon-assisted spectroscopy [12], [13]. Finally, a significant portion of the plasmonic energy can remain confined in the nanoparticle's immediate vicinity as intense, non-propagating evanescent fields. This near-field confinement enhances light–matter interactions on the nanometer scale, enabling sensing mechanisms based on local electromagnetic field amplification. Key sensing modalities that exploit this effect include surface-enhanced Raman scattering (SERS) [13], which allows ultrasensitive molecular fingerprinting down to single-molecule levels, plasmon-enhanced fluorescence (PEF) [14], which increases both the excitation and emission efficiency of fluorophores near the surface, and surface-enhanced infrared absorption (SEIRA) [15], which improves infrared molecular detection by boosting vibrational signatures.

These all-optical functionalities make plasmonic nanoparticles very interesting for the development of compact lab-on-fiber platforms [16]. Diverse plasmonic architectures have been realized on fiber tips to deliver optically controlled thermal stimuli at the microscale, including focused ion beam (FIB) milling [17], [18], electrostatic self-assembly [19] and deposition of porous plasmonic layers exploiting the non-radiative pathway for applications ranging from the accumulation of bacteria driven by localized thermo-plasmonic heating effect to the trapping of live cells and colloidal particles via thermophoresis and convection [20]. Some other relevant works have demonstrated fiber-based SERS platforms as highly sensitive, label-free chemical and biosensing substrates in both flat and tapered configurations for molecular fingerprinting [21], [22], [23], [24], [25]. Recently, several studies have demonstrated fiber probes integrating both functionalities—plasmonic heating and SERS—into a single platform [26], [27]. These probes rely on the same plasmonic nanostructures to perform localized heating via photothermal conversion and Raman signal enhancement, typically using a single excitation source. A near-infrared (NIR) laser (785 or 808 nm) is employed to simultaneously generate heat and drive SERS, with the photothermal effect arising from non-radiative plasmon decay at high excitation powers. While this approach successfully enables multifunctional fiber tools for cell manipulation, trapping, and biochemical detection, all reported configurations to date utilize a single wavelength to excite both functionalities. To the best of our knowledge, no study has yet demonstrated an optical fiber platform able to exploit two spectrally distinct excitation wavelengths— for instance one in the visible for controlled photothermal activation, and another in the NIR for SERS-based chemical interrogation. The realization of such dual-wavelength systems would offer complementary opportunities for sequential or decoupled thermal control and sensing, expanding the functional versatility of lab-on-fiber devices.

However, nanofabrication at the fiber tip of structures supporting dual-wavelength operation poses significant challenges, due to the small diameter and high aspect ratio of optical fibers [28], [29]. Top-down techniques, such as FIB milling or nanofabrication based on electron beam lithography have been already employed for fabricating dual band visible-NIR plasmonic systems [17]. Despite offering precise control over nanostructure geometry and several engineering degrees of freedom, when translated to optical fibers they suffer from long fabrication times, low scalability and generally low compatibility with nanofabrication protocols developed for flat substrates. Alternatively, electrostatic deposition methods offer better scalability and the ability to coat non-planar or curved surfaces, such as fiber tips [19] but often require complex, time-consuming protocols. An alternative method is represented by solid-state dewetting, which emerged as a simple, scalable, and effective bottom-up method to fabricate plasmonic nanostructures on fiber tips. In this process, thin gold films thermally reconfigure into gold nanoislands (AuNIs), whose morphology and density, and in turn their plasmonic characteristics, can be tuned by adjusting the initial film thickness [24] and, if desired, repeated dewetting cycles [23], [30]. These AuNIs exhibit strong optical absorption in the visible spectrum, making them effective for plasmonic heat generation, and provide considerable near-field enhancement in the near-infrared range for high-sensitivity SERS detection.

Building on this dual functionality we propose a multifunctional plasmonic fiber probe (PFP) that, to the best of our knowledge, is the first fiber-integrated device capable of switching between thermo-plasmonic heating in the visible and SERS sensing in the near-infrared by simply varying the excitation wavelength. The probe features AuNIs integrated onto a multimode optical fiber facet via solid-state dewetting. The resulting nanostructures possess tunable plasmonic properties that dictate the optical and thermal response of the device. We experimentally demonstrate that the thermo-plasmonic effects of the PFP depend strongly on the illumination wavelength and closely follow the probe's optical extinction spectrum. Comparative studies with gold nanofilms of similar thickness were conducted to highlight the heating efficiency of the dewetted AuNIs. To support our experimental findings, we developed comprehensive electromagnetic and thermal models using the finite element method (FEM). Numerical simulations were also carried out to evaluate the temperature gradients produced in liquid environments, critical for biomedical applications where direct temperature measurements are often impractical. We further demonstrate PFP as a SERS platform for detecting benzenethiol (BT) molecules under 785 nm excitation, leveraging the strong field enhancements provided by the AuNIs. These results underline the dual-mode capabilities of our fiber probe and demonstrate its potential as a powerful tool for biomedical diagnostics and sensing applications.

## 2 Results and Discussions

### 2.1 Device Fabrication and Optical Characteristics Analysis

The multifunctional plasmonic fiber probe (PFP) used in this study, illustrated in **Fig. 1(a)**, is based on a standard multimode silica optical fiber with a 200 μm core diameter. To functionalize the fiber tip with AuNIs, we employed a solid-state dewetting approach [22], [23]. As shown schematically in **Fig. 1(b)**, a thin gold film is first deposited onto the fiber facet via electron beam evaporation. The coated fibers are then subjected to thermal annealing: the temperature is ramped from room temperature to 600 °C at a rate of 10 °C/min, held at the target temperature for one hour, and subsequently cooled to ambient temperature under natural conditions (full details are provided in the Materials and Methods section). This fabrication method allows tuning of the AuNI size and density, parameters that directly influence the photothermal properties of the device, by adjusting the initial gold film thickness. For this work, we fabricated two variants of the multifunctional PFP: (i) PFP1, with an initial film thickness of 5 nm, and (ii) PFP2, with a film thickness of 10 nm. The resulting nanostructures are visualized in the scanning electron microscopy (SEM) images shown in **Fig. 1(c)**, which confirm a uniform distribution of AuNIs across the fiber facet for both variants. A statistical morphological analysis was performed using ImageJ software. The diameter distributions of the AuNIs were well-fitted by Gaussian profiles centered at 34 nm for PFP1 and 134 nm for PFP2. The surface coverage rates were found to be 36% for PFP1 and 28% for PFP2. Interparticle gaps increased with island size, measuring 16 nm for PFP1 and 92 nm for PFP2, respectively. A summary of these morphological parameters, including average diameter, interparticle spacing, and height, is provided in **Fig. 1(d)**.

To assess the optical characteristics of the probes, we measured the extinction spectra in transmission mode under broadband illumination through the fiber. This method is commonly used to evaluate light absorption by plasmonic nanoparticles [31]. The resulting spectra revealed resonance peaks at 540 nm for PFP1 and 570 nm for PFP2 (**Fig. 2(a)**). As expected, the larger AuNIs in PFP2 led to a redshift and broadening of the resonance peak, along with an overall increase in extinction magnitude. For comparison, we also characterized the optical extinction spectra of flat gold nanofilms deposited on the fiber facet. As shown in **Fig. 2(b)**, the extinction magnitude for the 5 nm gold nanofilm is lower than that of the 10 nm film up to around 600 nm. Beyond this point, the 5 nm film exhibits higher extinction. In contrast, the 10 nm film displays an almost flat extinction response across the visible spectrum. These observations are consistent with previously reported behavior of thermally evaporated gold films [32]. Further details on the experimental setup used for spectral measurements can be found in our earlier work [23] and Materials and Methods section.

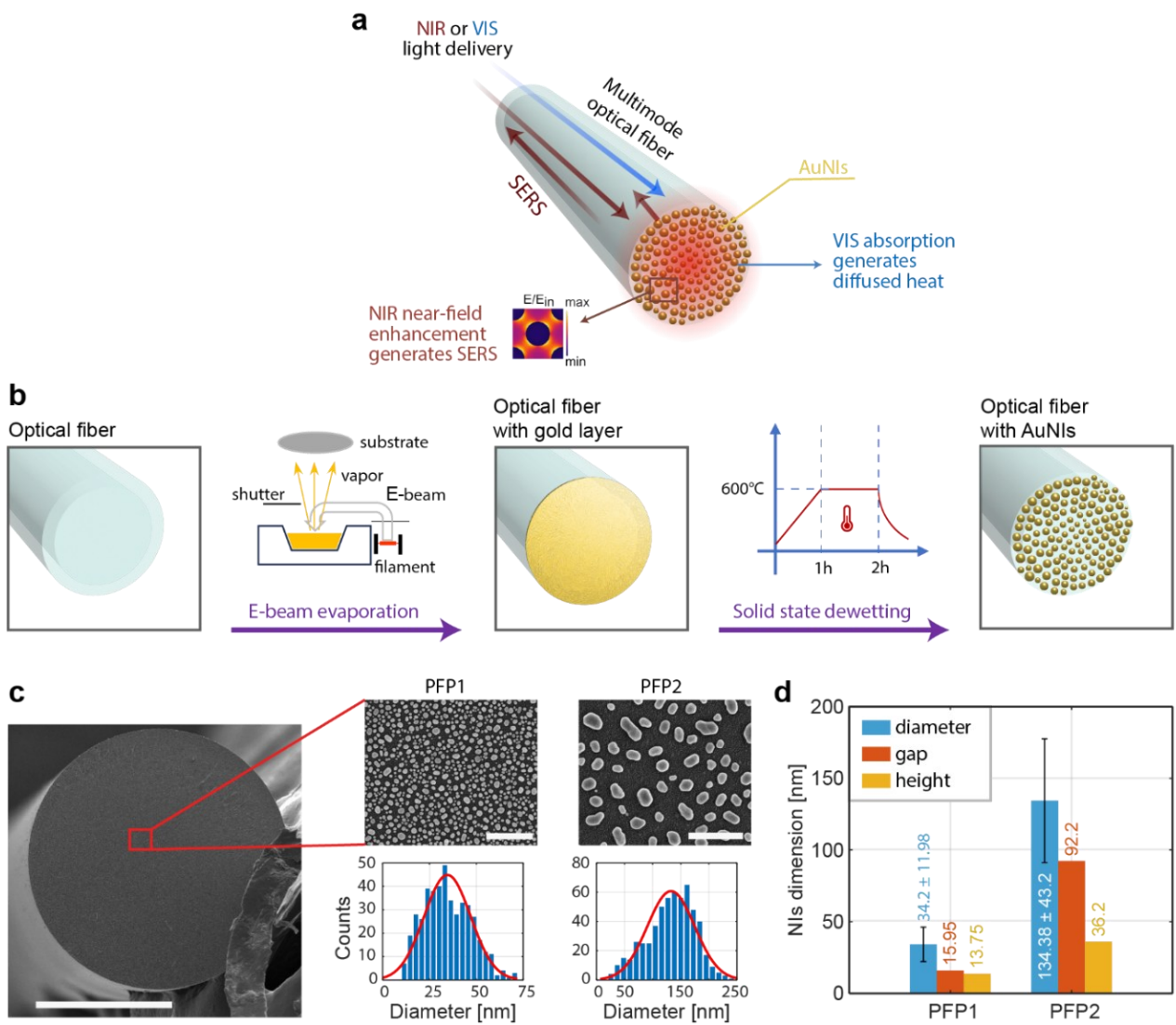

**Fig.1.** (a) A plasmonic optical fiber (PFP) schematic with AuNIs at the fiber facet. (b) Schematic illustration of the fabrication procedure. (c) SEM images showing the fiber facet (scale bar 100 µm) and the morphology of the AuNPs in the enlarged images (scale bar 500 nm) for both PFP1 and PFP2 along with the histograms of the AuNPs diameter fitted with a Gaussian distribution centered at 34 nm and 134 nm for PFP1 and PFP2, respectively. (d) Bar graphs of the statistical data of average diameter with standard deviation as error bar and effective gap as well as the height of the AuNIs of morphologies shown in (c).

To gain deeper insight into the optical behavior of the plasmonic fiber probes, particularly the absorbed power, which is directly responsible for photothermal effects, we conducted electromagnetic simulations using the finite element method (FEM) in COMSOL Multiphysics. In these simulations, the array of AuNIs was modeled as a periodic computational unit cell, where the nanoparticle diameter, height, and surface coverage were set to the average values extracted from our morphological analysis. Despite not representing in full the stochastic nature of the nanoparticles nucleation, this geometry has been previously validated to model the main plasmonic behavior of the system in our prior work [22], [23], [24]. Small variations in particle size and interparticle spacing were neglected, as they are known to have a minimal impact on the overall optical response, and using averaged parameters provides a reliable approximation for numerical modeling [33]. The lattice constant and effective interparticle gap were calculated, assuming a square lattice configuration in which each AuNI is surrounded by four neighboring islands positioned at the corners of a square, based on the average morphological parameters shown in **Fig. 1(d)**.

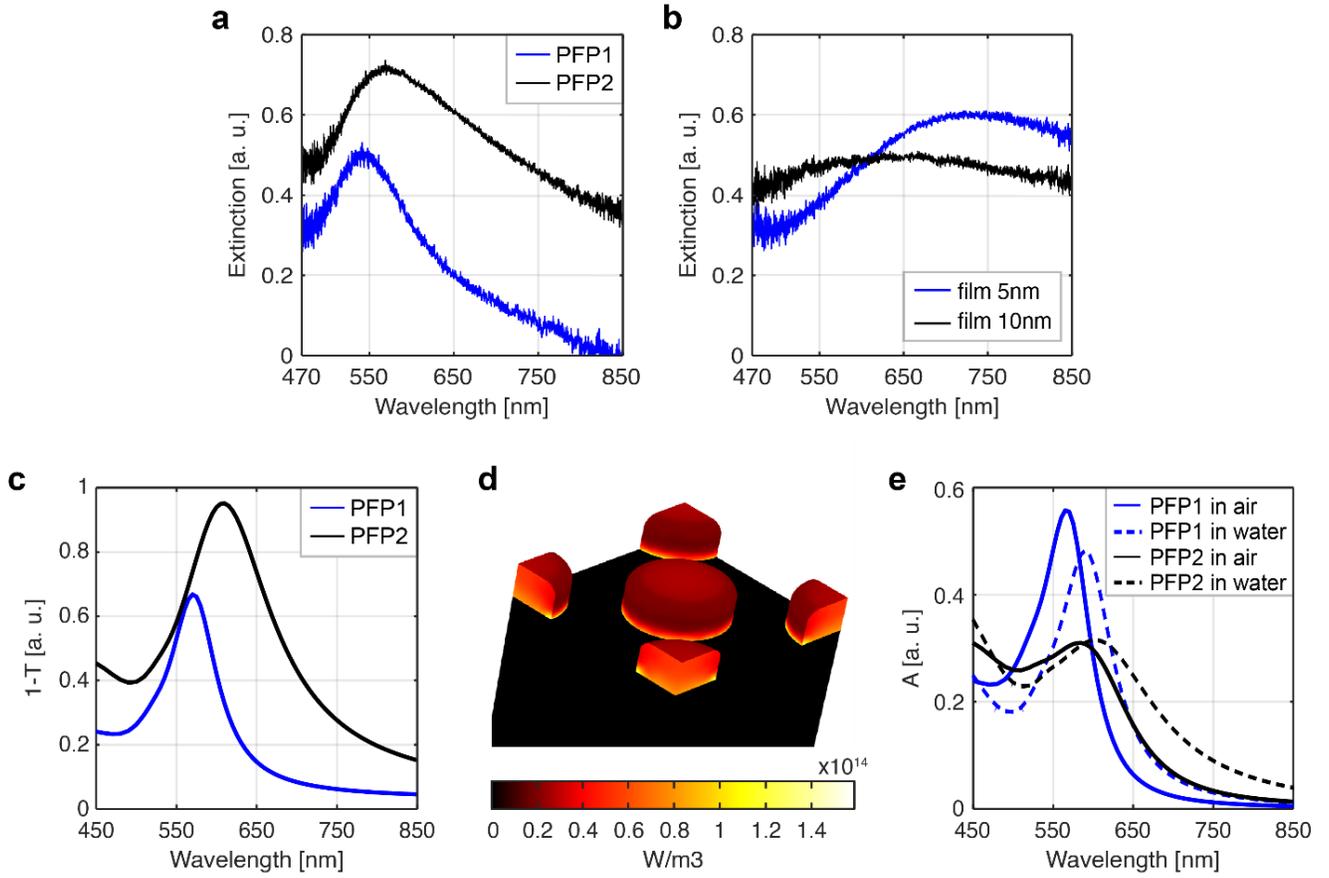

**Fig 2.** (a) The measured extinction spectra of the optical fiber probes taken in the transmission configuration showing resonance peaks at 540 nm (PFP1) and 570 nm (PFP2). (b) The measured extinction spectra of the optical fibers with 5 nm and 10 nm thick gold nanofilms at the fiber facet. (c) A plot of the computed 1-T spectra for both PFP1 and PFP2 is shown. (d) A color map of the volumetric absorbed power density computed at 473 nm for PFP1. (e) The absorbance (A) spectra for both PFP1 and PFP2 in air and water environments.

**Fig. 2(c)** shows the simulated $1 - T$ spectra (with $T$ representing transmittance), which were calculated to enable direct comparison with the measured extinction spectra presented in **Fig. 2(a)**. These simulations confirm that the observed optical resonances stem from the excitation of plasmonic modes in the AuNI arrays. The computed spectra replicate the experimental trends: for PFP2, the larger AuNIs and wider interparticle gaps produce a redshifted and broader resonance peak (FWHM = 160 nm) with a peak intensity of 0.95; in contrast, PFP1 exhibits a narrower, less intense peak (FWHM = 90 nm; amplitude = 0.66), due to the smaller particle size and higher density. Minor discrepancies in peak position and amplitude between simulation and measurement can be attributed to the random distribution and shape variability of the AuNIs in the actual devices.

For the thermo-plasmonic effect, the key quantity is the absorbed power density, which determines the amount of generated heat, for which, under optical excitation, the AuNIs absorb energy from the incident light and dissipate it as heat. The volumetric heat power density $q(r)$ scales with the square of the electric field and is expressed as [2]:

$$q(\mathbf{r}) = \frac{\omega}{2} \, Im(\varepsilon(\omega))\varepsilon_0 |\mathbf{E}(\mathbf{r})|^2, \tag{1}$$

where $\omega$ is the free space frequency of the light wave, $\varepsilon(\omega)$ is the complex dielectric constant of gold, $\varepsilon_0$ is the permittivity of the free space and $\mathbf{E}(\mathbf{r})$ is the electric field inside AuNIs. **Fig. 2(d)** shows the spatial distribution of the absorbed power density calculated over periodic unit cell. The total heat power resulting from the Joule heating effect is obtained by integrating $q(\boldsymbol{r})$ over the volume of the nanoparticles using the equation (2).

$$P_{abs} = \int_V q(\boldsymbol{r}) \, d^3r \tag{2}$$

The absorbance spectra are shown in **Fig. 2(e)** for both PFP1 and PFP2, under air and water background conditions. The position, width, and intensity of the absorption peaks depend strongly on the interparticle spacing (i.e., the array periodicity). As the interparticle gap increases, the resonance peak redshifts, broadens (higher FWHM), and diminishes in intensity [34]. This explains why PFP1, with denser AuNI packing and smaller gaps, shows a stronger absorption peak compared to PFP2.

## 2.2 Heating Characteristics Analysis

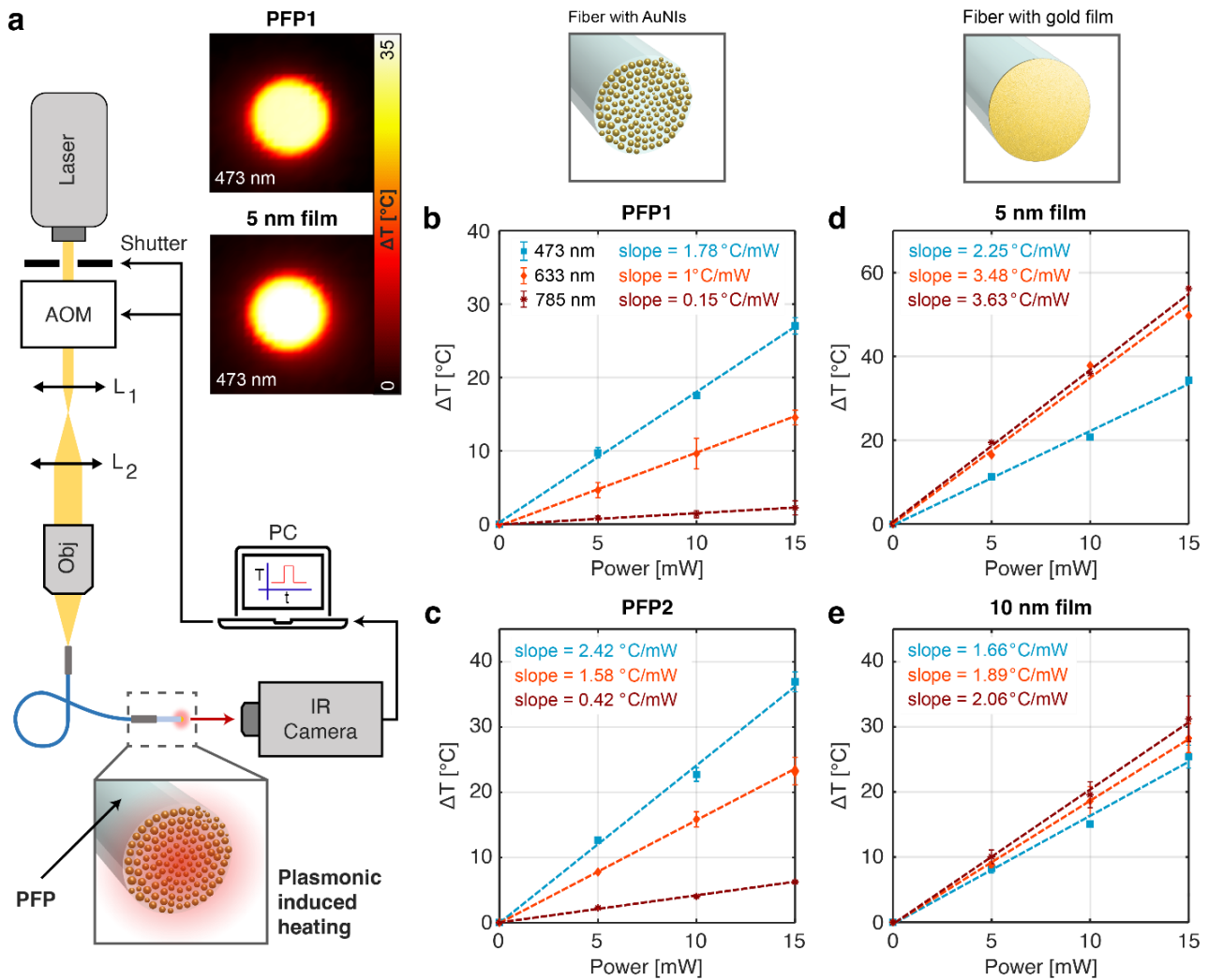

**Fig 3.** Results of temperature measurements with thermographic camera in air. (a) A schematic representation of the measurement setup. The insets show the measured heat map of the fiber facet. The images show an area of 20×20 pixels (488×488 µm²). (b-c) The photothermal efficiency produced by PFP1 and PFP2 in air as the laser power is linearly increased at three wavelengths of 473 nm, 633 nm, and 785 nm. (d-e) The photothermal efficiency produced in air by 5 nm and 10 nm thick gold films at the fiber facet as the laser power is linearly increased at three wavelengths of 473 nm, 633 nm, and 785 nm.

To characterize the temperature increase at the fiber facet induced by plasmonic heating, we performed thermographic measurements using a high-resolution infrared thermal camera. The details of the experimental setup are provided in the Materials and Methods section. The results of steady-state temperature measurements are presented in **Fig. 3**. The excitation laser beam was collimated and directed through an objective lens, to be then coupled inside the PFP, as schematically shown in **Fig. 3(a)**. We used three continuous-wave (CW) lasers with wavelengths of 473 nm, 633 nm, and 785 nm, covering the visible to near-infrared range, relevant for biomedical applications. Laser power was modulated using an acousto-optic modulator (AOM). The insets in **Fig. 3(a)** show representative thermal maps of the fiber facet. Due to the high density and close proximity of the AuNIs, the temperature distribution appears smooth and uniform, as result of collective photothermal heating from neighboring nanoparticles [7], [35].

The maximum steady-state temperature increases for PFP1 and PFP2, measured as a function of increasing laser power, are shown in **Figs. 3(b–c)**. Error bars represent the standard deviation, and the dashed lines are fitted curves. For comparison, **Figs. 3(d–e)** display the temperature rise measured for 5 nm and 10 nm thick gold nanofilms deposited on the fiber facet under the same conditions. The photothermal efficiency (PE) is defined as the maximum temperature increase per milliwatt of injected laser power [19]. The measured PE values for PFP1 and PFP2 correlate well with both their experimental and simulated extinction spectra (**Fig. 2(a)**). For both devices, 473 nm and 633 nm wavelengths yield higher PE than 785 nm, due to stronger light absorption near the plasmonic resonance. The higher PE of PFP2 compared to PFP1 aligns with its greater optical absorbance across all tested wavelengths. The 5 nm gold nanofilm exhibited higher PE than the 10 nm film at all three wavelengths, despite having similar extinction properties. This discrepancy is attributed to more efficient heat dissipation in thicker films, which spreads the absorbed energy over a larger volume and results in a lower surface temperature [36]. Additionally, as described in [32], the higher thermal conductivity of the 10 nm gold film acts as a heat sink, further reducing the maximum achievable temperature. When comparing the PFP devices with gold nanofilms, it is evident that the PFPs exhibit plasmonic heating behavior, with maximum heating at 473 nm and minimum at 785 nm, closely following the extinction spectra. In contrast, gold nanofilms produce more uniform heating across the visible–NIR spectrum, showing an opposite trend. These findings confirm that the plasmonic heating of PFPs is highly tunable via the excitation wavelength and the morphology of the AuNIs. While gold nanofilms also generate heat and are used in photothermal applications such as neural interfaces [32], their broad spectral response lacks the sharp resonance behavior of plasmonic structures.

To further analyze the spatial and temporal characteristics of heat transfer, we developed a numerical model using the Heat Transfer Module in COMSOL Multiphysics. This model solves the time-dependent heat diffusion equation (3) in a domain with arbitrary geometry:

$$\rho C_p \frac{\partial T}{\partial t} = \kappa \nabla^2 T + q(r), \tag{3}$$

where $\rho$ is the density (kg/m$^3$), $C_p$ is the specific heat capacity (J/kg.K), $\kappa$ is the thermal conductivity (W/m.K) of the materials involved, and $q(r)$ is the volumetric heat power density (W/m$^3$) induced by the AuNIs. All material parameters used in our models are listed in Table 2 in the Materials and Methods section.

Given the nanoscale interparticle gaps and dense packing of the AuNIs, we followed the modeling approach described in [37], where the collective heating of closely spaced particles is approximated as a single effective heat source. Accordingly, we implemented a thin gold film as an equivalent boundary heat source, using the heat power from the FEM simulations as input heat power rate. The model geometry is shown in **Fig. 4(a)**, with a large spherical domain representing the surrounding environment (air in this case). The simulated temperature map in **Fig. 4(b)** shows the spatial temperature distribution near the fiber tip. The inset highlights the steady-state temperature profile in the xy-plane at the fiber facet. **Fig. 4(c)** illustrates the temperature decay profile, showing that the thermal gradient drops to 20% of its maximum value within 500 μm from the fiber and returns to ambient temperature at the 10 mm boundary of the simulation domain. Both simulated and measured PEs display a linear relationship with the input power, consistent with the linearity of the heat diffusion equation (Eq. 2) that governs the plasmonic heating phenomenon. Comparing the simulated PEs in **Figs. 4(d–e)** with the experimental data in **Figs. 3(b–c)** reveals that the simulations slightly overestimate the temperature rise. This overestimation arises from modeling simplifications, such as using a periodic unit cell in the electromagnetic solver. Another reason can be variation in the values of physical parameters (e.g., thermal conductivity and heat capacity). Moreover, the thermal camera's limited spatial resolution averages the temperature over several pixels, leading to a lower measured peak temperature than the actual maximum. Despite these factors, the numerical predictions remain in reasonably good agreement with the experimental data, validating the overall modeling approach.

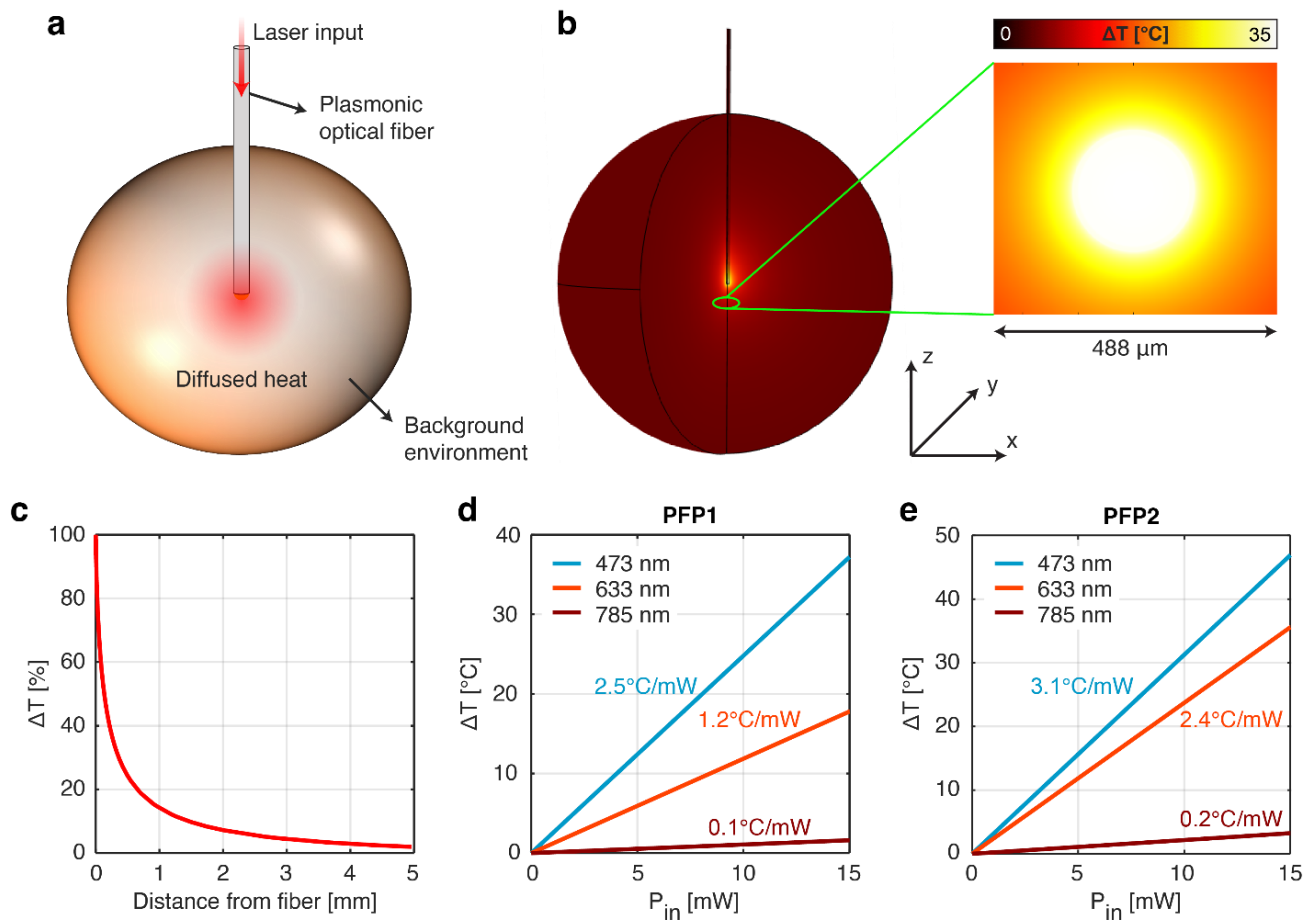

**Fig 4.** Results of steady-state temperature calculations. (a) A schematic representation of the simulation domain. (b) A 3D color map showing the spatial temperature profile of the PFP under laser (473 nm) excitation at 15 mW power. The inset shows the color map of the temperature gradient at the PFP facet in the xy plane. (c) Spatial temperature distribution as a function of the distance from the PFP. The panels (d) and (e) show the calculated photothermal efficiency of PFP1 and PFP2 in the air at wavelengths of 473 nm, 633 nm, and 785 nm when laser power is linearly increased from 0 to 15 mW.

To investigate the thermo-plasmonic behavior of the PFP under *in vitro* conditions for perspective biological applications, we employed numerical simulations to model the spatial and temporal heat transfer characteristics with water as the surrounding medium. Compared to air, water's higher thermal conductivity leads to a significant reduction in temperature gradients generated at the fiber tip. The simulated PEs of PFP1 and PFP2 in water are presented in **Figs. 5(a-b)**. The results show that for PFP1, an input power of 20 mW at 473 nm produces a temperature increase of 5.1 °C (PE=0. 26°C/mW), while the same power at 633 nm results in a 6.38 °C temperature gradient (PE=0.38°C/mW). The higher PE at 633 nm aligns with the absorbance spectrum of PFP1 in water, as shown in **Fig. 2(e)**. However, at 785 nm, the PE drops substantially, yielding only 0.5 °C of temperature increase at the same power level (PE=0.03°C/mW). For PFP2, under 20 mW laser excitation, the temperature increases at the fiber tip are 7.66 °C at 473 nm (PE=0.38°C/mW), 7.2 °C at 633 nm (PE=0. 26°C/mW), and 1.66 °C at 785 nm (PE=0.08°C/mW). On average, the simulations indicate a ~90% reduction in PE at 473 nm and ~80% at 633 nm for both PFP1 and PFP2 when transitioning from air to water. At 785 nm, the generated temperature increase remains below 2 °C even for 20mW input, making it negligible for most thermal stimulation applications.

The spatial temperature distribution in water under 20 mW excitation at 473 nm is shown in **Fig. 5(c)**. The heat map reveals the localized nature of the heating effect, while the line plot quantifies the temperature variation as a function of radial distance from the fiber tip. The temperature decays to 20% of its peak value within 360 μm, demonstrating a more localized heating profile than in air. Interestingly, although the peak temperature is lower in water, the local confinement of heat is stronger (by ~28%) due to water's higher heat capacity, which prevents rapid heat diffusion away from the fiber tip. We also investigated the thermal transient response of PFPs in water. Time-dependent simulations of the temperature gradients for PFP1 and PFP2 generated at varying input powers (5 mW to 20 mW) are shown in **Figs. 5(d-e)**. The thermal time constant, defined as the time required to reach 90% of the steady-state temperature, was found to be approximately 650 ms, averaged over both PFP1 and PFP2. Notably, the transient response is independent of input power. These results are consistent with previous reports on fiber-based thermo-plasmonic heaters employing gold nanorods as the heat-generating elements [19], confirming that while plasmonic heating remains effective in liquids, the reduced temperature gradients and slower heat dissipation must be considered when designing applications for *in vitro* or *in vivo* environments.

## 2.3 Through fiber SERS measurement

The plasmonic response of the AuNIs was confirmed by the extinction spectra shown in **Fig. 2(a)**. These spectra reveal strong absorption in the visible range, peaking around the plasmonic resonance, and progressively decreasing toward the infrared. This behavior enables efficient thermo-plasmonic heating at shorter wavelengths (e.g., 473 nm), while still maintaining sufficient plasmonic activity at 785 nm to support surface-enhanced Raman scattering (SERS). Therefore, by capitalizing on both absorption and near-field enhancement capabilities of AuNIs, the same PFP serves a dual functionality: plasmonic heating under visible excitation, and SERS-based molecular detection at 785 nm, a commonly used wavelength for Raman spectroscopy in biological systems by virtue of its low fluorescence background. To evaluate the SERS performance, we performed electromagnetic simulations that show the near-field enhancement remains considerable at 785 nm, particularly in the nanoscale gaps between AuNIs. The field enhancement versus wavelength curves for both PFP1 and PFP2 are presented in **Fig. 6(a)**, with insets illustrating the distributions of near-field enhancement at 785 nm. These results show that both PFP geometries can support SERS activity at this excitation wavelength.

The SERS performance of the PFPs were characterized with benzenethiol (BT) molecules functionalized on the AuNIs surface with a custom-built bench-top Raman microscope (details in Methods). The resulting SERS spectra collected with PFP1 and PFP2 in a through-fiber configration with 785 nm excitation are shown in **Figs. 6(b)** and **6(c)**, respectively. As the BT signature peaks were hardly observed from blank and thin film covered fiber [22], the enhancment of the BT signartures peaks were obvious in the PFP device. The peaks were observed at 999, 1025, 1069, and 1574 cm$^{-1}$, corresponding to well-known vibrational modes of the BT molecule. To isolate the SERS contribution, the silica background signal, measured using a bare PFP (without BT coating), was calibrated to 802 cm$^{-1}$ silica peak in the SERS spectra and then substracted. To assess the dependence of the SERS enhancement on the input power, additional measurements were conducted by varying the in-fiber excitation power. The results are shown in **Fig. 6(d)**. Notably, SERS enhancement was detectable at power levels as low as 1 mW, demonstrating the high sensitivity of the platform even at low excitation powers.

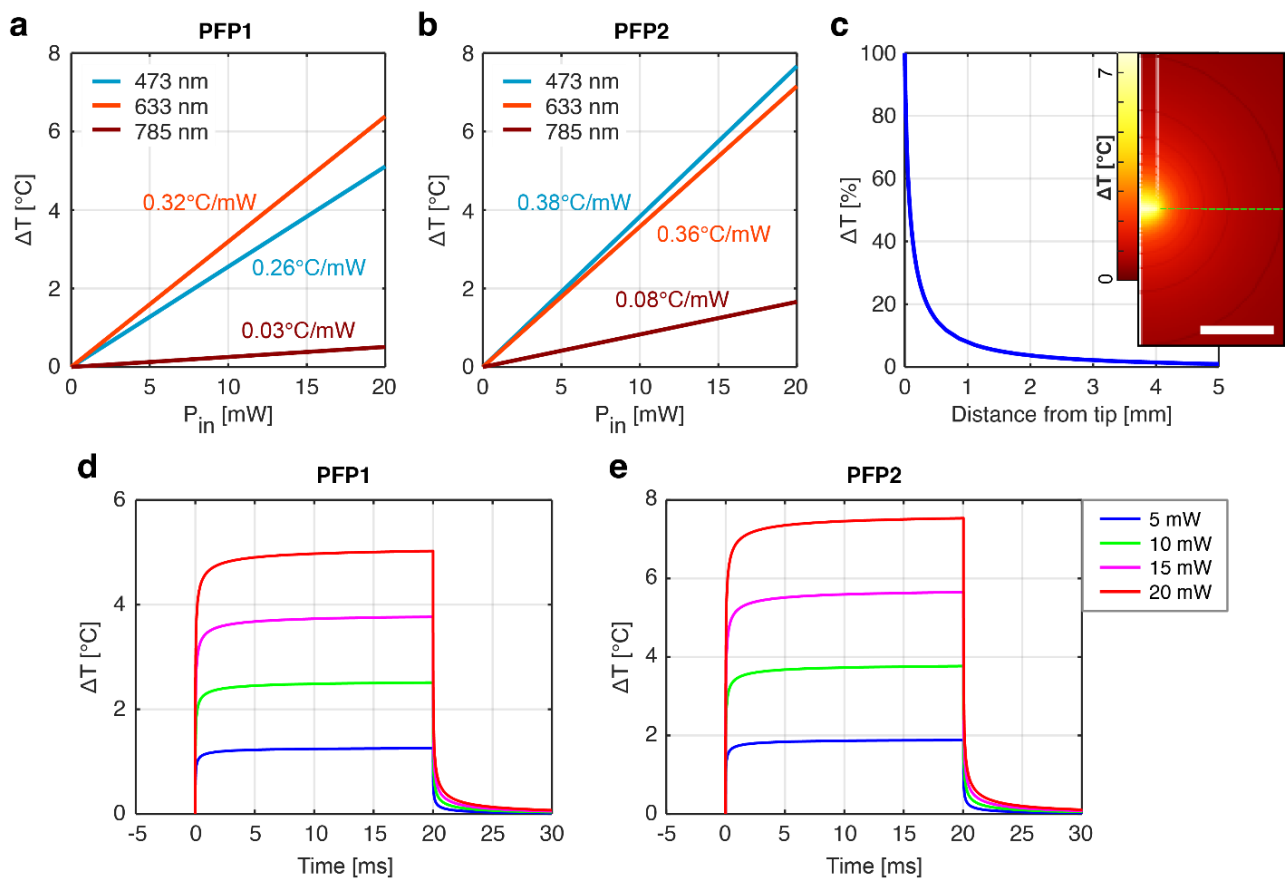

**Fig.5.** Results of spatial and temporal temperature calculation in water. The panels (a) and (b) show the calculated photothermal efficiency of PFP1 and PFP2 in water at wavelengths of 473 nm, 633 nm, and 785 nm when laser power is linearly increased from 0 to 20 mW. (c) The temperature profile as a function of the distance from the PFP tip. The color map shows the spatial temperature distribution in the water around the PFP tip under a laser power of 20 mW at 473 nm. The temperature gradient is plotted on the dotted green line shown in the color map. (d-e) The simulated transient responses of the PFP1 and PFP2 are shown.

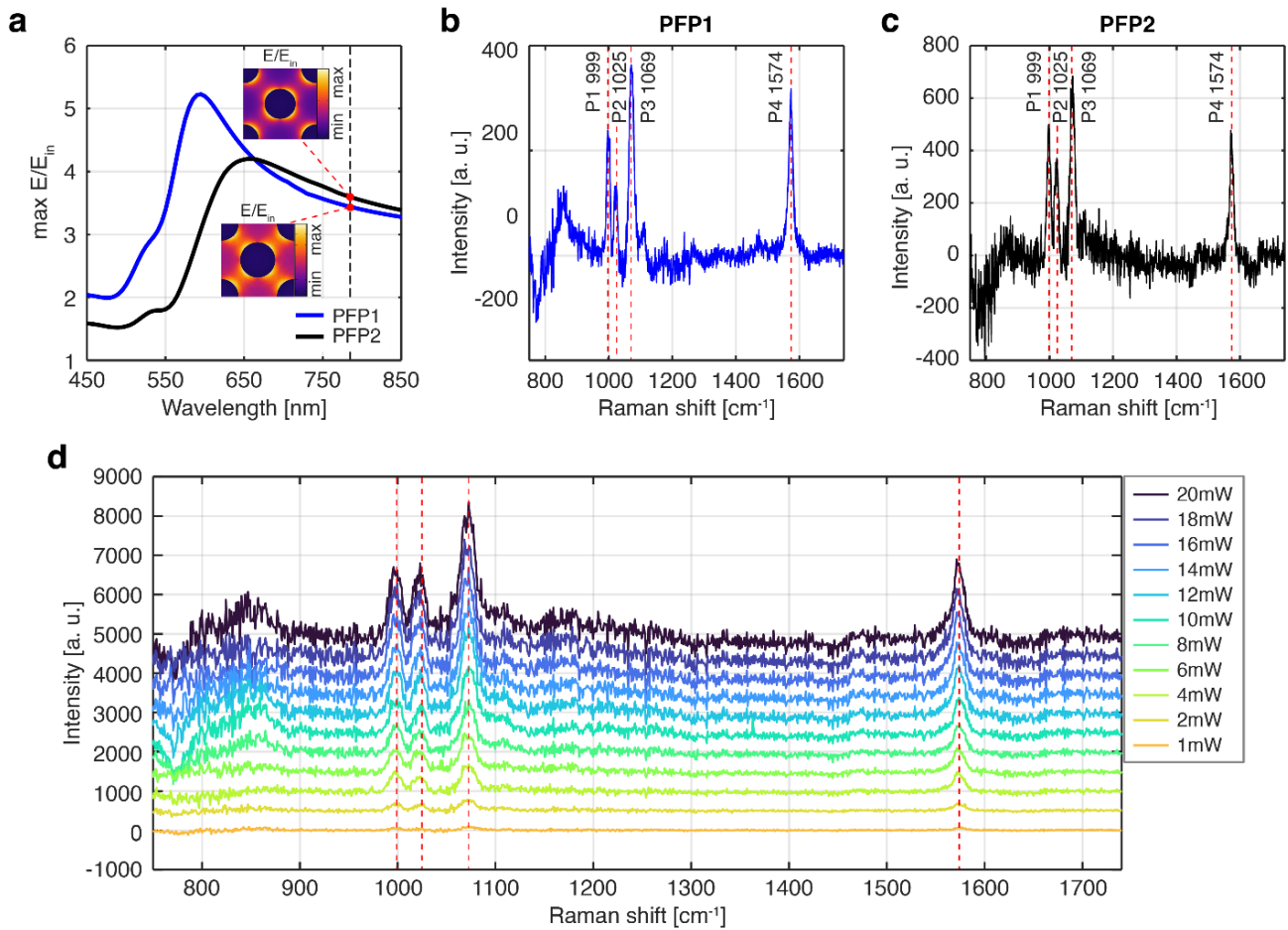

**Fig. 6.** Through fiber SERS response of BT molecules with the PFP at 785 nm as function of the input power at the fiber facet. (a) The simulated maximum field-enhancement induced by AuNIs assembly (for both PFP1 and PFP2) as a function of wavelengths is shown. The inset shows the color map of the filed profile at 785 nm. (c-d) SERS measurements taken by PFP1(blue) and PFP2 (black) with 785 nm excitation (4 mW input power). The vertical lines indicate the molecular signature peaks at 999, 1025, 1069 and 1574 cm$^{-1}$. (d) SERS measurements taken by PFP2 as a function of the input power. All the SERS spectra shown here are background subtracted and normalized to the silica peak at 802 cm$^{-1}$.

## 3  Conclusion

In this work, we have presented the design, fabrication, and characterization of a multifunctional plasmonic fiber probe (PFP) operating at two different wavelengths to generate plasmonic heat and SERS signal, thoroughly analyzing its optical and thermal properties through both experimental measurements and numerical simulations. The probe is based on a simple and reproducible fabrication approach that employs solid-state dewetting to form gold nanoislands (AuNIs) on the flat facet of a silica optical fiber. This technique eliminates the need for complex nanoparticle synthesis or non-standard coating procedures and enables the formation of thermally stable nanoparticles.

The morphology of the AuNIs is controlled by tuning the initial gold film thickness during evaporation, offering a route to optimize both the optical response and the thermo-plasmonic behavior of the probe. The resulting AuNIs exhibit broadband optical activity, with higher absorption in the visible and substantially preserve near-field enhancement in the near-infrared. Specifically, the strong absorption in the visible range enables efficient thermo-plasmonic heating at excitation wavelengths such as 473 nm, while the field enhancement at 785 nm support the use of the same probe as an active SERS platform. These findings demonstrate that a single AuNI-based probe based on multimodal optical fiber can operate in a dual-functional mode: (i) Thermo-plasmonic heating under visible excitation (473 nm), and (ii) Surface-enhanced Raman scattering (SERS) sensing under near-infrared excitation (785 nm).

We also demonstrated that the heating characteristics of the probe can be tuned by adjusting the AuNI morphology. Increasing the initial film thickness from 5 nm to 10 nm results in a change of AuNIs morphology and higher optical extinction and greater photothermal efficiency (PE) due to enhanced plasmonic absorption. The PFPs exhibit a linear temperature increase at the fiber facet in response to input optical power, consistent with predictions from our finite element method (FEM) simulations. Numerical modeling further revealed that, in a liquid environment, temperature gradients of 5-8 °C can be achieved using 20 mW input power at 473 nm or 633 nm. The heating is spatially localized, with approximately 50% of the thermal gradient confined within an 85 μm radius around the fiber tip. The probes also exhibit a transient thermal response time of ~650 ms, in line with values reported for comparable photothermal systems.

Taken together, our experimental and simulation results underscore the potential of this multifunctional optical fiber platform for a wide range of biomedical applications. The inherent advantages of optical fibers, such as miniaturization, remote light delivery, and in vivo compatibility, combined with the dual capabilities of thermo-plasmonic actuation and SERS-based detection, position this system as a powerful tool for integrated diagnostics and therapy. For instance, thermo-plasmonic functionality enables controlled photothermal effects that can be utilized for applications such as targeted hyperthermia, drug delivery, or tissue ablation, while the SERS functionality could provide label-free detection of biomolecules. Looking forward, such multifunctional fiber probes are expected to play a key role in the development of lab-on-fiber systems, where diagnostic sensing and therapeutic functions are integrated into a single, compact platform suitable for real-time, in situ biomedical use.

## 4 Material and Methods

### 4.1 Device fabrication and optical characterization

Standard multimode silica optical fibers (FG200LEA, 0.22 NA, Low-OH, Ø200 μm Core) were used for the device fabrication. As a first step, the acrylate jacket was mechanically removed after immersing the fibers (7.5 cm long) in an acetone bath for 30 min. Then the fibers were cut on one side with a manual fiber cleaver (Thorlabs, XL411) to obtain a flat fiber facet. As a next step, a thin gold film was deposited on the fiber facet using an e-beam evaporator (Thermionics laboratory, inc. e-Gun™). The fibers were held with a 3D-printed custom-built mount to align the PFP surfaces of all the fibers normal to the gold source. During the evaporation process, a deposition rate of 0.2 Å s$^{-1}$ under a chamber pressure of $6 \times 10^{-6}$ mbar was applied. Finally, the fibers were transferred to a muffle furnace (Nabertherm B180) for the thermal annealing step. The fibers were placed in a ceramic bowl. The furnace temperature was gradually increased from room temperature to 600 °C at a rate of 10 °C/min and was held constant at 600 °C for an hour. After that, it was allowed to be cooled down to room temperature at ambient conditions. This resulted in the formation of AuNIs at the fiber facet. Eventually, metallic ferrules (1.25 mm) were connected on the other end of the fibers and manually polished for optical and thermal characterization.

### 4.2 Thermal characterization

A high-resolution thermographic infrared camera (FLIR A600-series) was used to measure the temperature at the facet of PFP in air as a background environment. The thermal camera has a focal plane array (FPA), an uncooled microbolometer, that works in the 7.5-14 µm spectral range and has an IR resolution of 640×480 pixels at 25 Hz frame rate. We used a close-up lens with a working distance of 46 mm. This setup yields a single-pixel resolution of 24.4 µm. Data acquisition was done through the FLIR ResearchIR software provided by the camera manufacturer.

As schematically shown in **Fig.3(a)**, the laser power was injected into PFP by collimating and expanding the laser beam and sending it to the entrance of the objective lens. Continuous wave laser sources of 473 nm, 633 nm, and 785 nm were used. The power was linearly increased from 0 to 15 mW. This corresponds to the power measured at the input of the PFP. An acousto-optic modulator (AA Opto-Electronic MT80-A1.5-VIS) was used to control the laser power. The room temperature was taken as the starting temperature. After the laser is turned on, the PFP facet heats up immediately, and the thermal image is taken by the thermographic camera after 1-2 min once the temperature reaches saturation for the given input power i.e., a steady state. Two devices of each type were used for temperature measurements. For each device under test, the temperature readings were taken twice at each power level, once when the power level was increased and second when the power level was decreased. The average of two readings was used. For the transient response, the laser power injected into the fiber is controlled by a shutter. The shutter is opened at time t = 0 s. After around (approximately 10 s), the shutter is turned off to block the laser power.

### 4.3 Optical characterization

For the extinction spectra measurement, a blank fiber was used for the subtraction and normalization of the measured extinction spectra of the PFP fibers following the relation $Extc_{AuNIs} = (T_{blank} - T_{AuNIs})/T_{blank}$, the experimental configration can be found in our previous work [23]. For the SERS measurement, home-built Raman microscope was used to characterized the PFPs in the through-fiber configuration, the details of the setup is in our previous publication [23], before the SERS measurement, the facet of the fabricated PFPs were immersed in BT molecule methanol solution (6 mM) for 3 h, the fibers were then rinsed by stirring them in a beaker of clean methanol for 10 minutes. This rinsing process was repeated three times with fresh methanol each time to ensure successful monolayer functionalization. Benzenethiol (BT, C6H5SH) molecules were purchased from Sigma-Aldrich (99%). After the BT functionalization, the SERS spectra were measured with an acquisition time of 60 s for all under the 785 nm continue laser excitaiton. The excitation power ranged from 1 to 20 mW. For all SERS spectra, the silica background, measured from a PFP without BT coating, was subtracted after calibration to the 802 cm$^{-1}$ silica peak.

### 4.4 Numerical Simulations

Electromagnetic and thermal simulations were carried out using the commercial FEM-based software COMSOL Multiphysics. For EM simulations, the distribution of AuNIs at the fiber facet was considered to be a square lattice [38] having particles at the center as well as at the corners of the square unit cell as illustrated in Fig.2(a). The effective interparticle gaps were calculated based on the period. The values of the period of the unit cell were derived from the knowledge of coverage rate and considering a square lattice for both PFP1 and PFP2 following the relation, $coverage\ rate = 2\pi(\frac{D}{2})^2/period^2$, where D is the average diameter of the AuNIs. For our devices, the period and the effective gaps were found to be 71 nm (16 nm) and 320 nm (92 nm) for PFP1 (PFP2). The periodic boundary conditions were implemented on the four side walls of the computational unit cell. The unpolarized plane wave was normally incident on the AuNIs array from the substrate. The full-wave solution of the electromagnetic fields was used. Perfectly matched layer boundary conditions were implemented on the PFP and the bottom surfaces to completely absorb the propagating waves. The AuNIs were modeled as cylinders with rounded corners to mimic their droplet-like shape. A refractive index of 1.4 for glass substrate was assumed. For gold, the optical constants were taken from [39].

Thermal simulations were performed using the heat transfer module. The geometry of the background environment in our model was set as a large sphere having a diameter of 10 mm. It should be large enough to diminish the boundary effects on the temperature. The temperature of the outer boundaries of the spherical domain was fixed to match the environment i.e., 20 °C for temperature calculations in the air background environment. The optical fiber was modeled as a glass cylinder having a diameter of 220 μm. A thin cylinder of gold having a diameter of 200 μm and a height of 10 nm was used as a heat source. The absorbed power calculated by electromagnetic simulations for a unit cell and scaled to the entire fiber facet was employed to the heat source. Since an optical fiber coated with a thin gold film is axially symmetric, a two-dimensional (2D) model of the optical fiber probe was implemented. The values of the material properties used in our model for the different materials involved (i.e. air, glass, gold, water) are given in Table 2.

**Table 2.** Details of the material properties used in our simulations [19].

| Property | Material | | | |
| --- | --- | --- | --- | --- |
|  | Gold | Glass | Air | Water |
| Thermal conductivity (W/m.K) | 314 | 1.38 | 0.02617 | 0.623 |
| Density (kg/m$^3$) | 19320 | 2203 | 1.118 | 993.37 |
| Specific heat (J/kg.K) | 125.604 | 730 | 1006 | 4180 |

# 5 Data Availability Statement

The data supporting the findings of this study are currently being curated and will be made available in a public repository upon publication of the article. Reasonable requests for access to the data prior to publication can be directed to the corresponding author.

# 6 Conflict of Interest Disclosure

The authors declare no conflict of interest.

# 7 Acknowledgments

M.F.K., D.Z., L.C., F.T., F.D.A, M.D.V, F.P. acknowledge funding from the European Union's Horizon 2020 research and innovation program under a grant agreement (# 828972). M.D.V., F.P. Acknowledge funding by the European Union (ERC, MINING, 101125498). Views and opinions expressed are however those of the author(s) only and do not necessarily reflect those of the European Union or the European Research Council. Neither the European Union nor the granting authority can be held responsible for them. M.D.V. and F.P. acknowledge that this project has received funding from the European Union's Horizon 2020 Research and Innovation Program under Grant Agreement No. 101016787. M.F.K and D.R. acknowledge funding under the Italian National Recovery and Resilience Plan (NRRP) of European Union – NextGenerationEU, partnership on "Telecommunications of the Future" (PE00000001 - program "RESTART"). A.B. acknowledges funding from the

European Union's Horizon 2020 research and innovation programme under the Marie Sklodowska-Curie grant agreement (# 101106602). FP and MDV acknowledge financial support under the National Recovery and Resilience Plan (NRRP), Mission 4 "Education and Research" – Component 2 "From Research to Business" – Investment 1.3, funded by the European Union – NextGenerationEU, within the framework of the cascade funding call of Spoke 3 – Extended Partnerships among Universities, Research Centers, and Enterprises for Research Project Funding, "Mnesys" (PE00000006). This work is part of the project "Correlating multifunctional signals in 3D printed neurochips for next generation studies on neural degeneration, via spectroscopy and electrophysiological recordings" (acronym: COMBO) – CUP: E63C22002170007